\documentclass[12pt]{article}  
\usepackage{graphicx}
\def\sq{\hbox {\rlap{$\sqcap$}$\sqcup$}}
\def\sq{\hbox {\rlap{$\sqcap$}$\sqcup$}}
\def\R{ {\rm R \kern -.31cm I \kern .15cm}}
\def\C{ {\rm C \kern -.15cm \vrule width.5pt \kern .12cm}}
\def\Z{ {\rm Z \kern -.27cm \angle \kern .02cm}}
\def\N{ {\rm N \kern -.26cm \vrule width.4pt \kern .10cm}}
\def\1{{\rm 1\mskip-4.5mu l} }
\def\lsim{\raise0.3ex\hbox{$<$\kern-0.75em\raise-1.1ex\hbox{$\sim$}}}
\def\gsim{\raise0.3ex\hbox{$>$\kern-0.75em\raise-1.1ex\hbox{$\sim$}}}
\def\noi{\noindent}

\def\beq{\begin{equation}}   \def\eeq{\end{equation}}
\def\bea{\begin{eqnarray}}  \def\eea{\end{eqnarray}}
\def\nn{\nonumber}
\def\noi{\noindent}
\def\beeq{\begin{eqnarray}} \def\eeeq{\end{eqnarray}}

\begin{document}

\begin{center} {\large \bf } \vskip 2 truemm

 {\large \bf Potentials for which the Radial} \vskip 2 truemm
 {\large \bf Schr\"odinger Equation can be solved}

 \par \vskip 5 truemm

{\bf Khosrow Chadan }\\ {\it Laboratoire de Physique
Th\'eorique}\footnote{Unit\'e Mixte de Recherche UMR 8627 - CNRS}\\   
{\it Universit\'e de Paris XI, B\^atiment 210, 91405 Orsay Cedex,
France}\\
{\it  (Khosrow.Chadan@th.u-psud.fr)} \par \vskip 5 truemm
{and}\par \vskip 5 truemm 
{\bf Reido Kobayashi} \\ 
{\it
Department of Mathematics}\\ {\it  Tokyo University of Sciences, Noda,
Chiba 278-8510, Japan}\\
{\it (reido@ma.noda.tus.ac.jp)}
\end{center}
\vskip 1 truecm

\begin{abstract}
In a previous paper$^1$, submitted to Journal of Physics A -- we presented an infinite class of potentials for which the radial Schr\"odinger equation at zero energy can be solved explicitely. For part of them, the angular momentum must be zero, but for the other part (also infinite), one can have any angular momentum. In the present paper, we study a simple subclass (also infinite) of the whole class for which the solution of the Schr\"odinger equation is simpler than in the general case. This subclass is obtained by combining another approach together with the general approach of the previous paper. Once this is achieved, one can then see that one can in fact combine the two approaches in full generality, and obtain a much larger class of potentials than the class found in ref. $^1$ We mention here that our results are explicit, and when exhibited, one can check in a straightforward manner their validity.\end{abstract}

\vskip 3 truecm 

\begin{flushleft}
LPT Orsay  01-2006\\
January 2006
\end{flushleft}
 
\newpage
\pagestyle{plain}
\baselineskip 20pt
\noi {\bf I -  Introduction} \\

In a recent paper,$^1$ we showed how, starting from two regular potentials for which the radial Schr\"odinger equation can be solved explicitly at zero energy --there are many of them$^{1,2}$-- one can construct explicitely an infinite number of potentials for which one can solve again, explicitely, the radial equation at zero energy. The solutions for these new potentials are given very simply in terms of the solutions  of the two initial potentials. By construction, it is then seen that the method can be applied without any modifications to potentials which are singular (repulsive) at the origin, or are long range (Coulomb, etc), or are even confining, like $\lambda r^2$, $\lambda > 0$, etc. According to the case, one can include, as well, the angular momentum potential $\ell (\ell + 1)/r^2$. Many examples, both for regular and singular potentials, covering all cases, except the confining potentials, are given in ref. $^1$, with explicit solutions of the radial equation. 

In the present paper, following a different method, we study a subclass of the general class, which is simpler, and more explicit. As in the general case, one can have, here too, singular (repulsive) potentials at the origin, or long range potentials like Coulomb potential, etc.

For the convenience of the reader, we give here a r\'esum\'e of ref. $^1$, where all the proofs can be found, together with appropriate references, which are essentially those of the present paper$^{2-8}$.

Consider first the radial Schr\"odinger equation at zero energy for the $S$-wave

$$\left \{ \begin{array}{l} \varphi ''_0 (r) = V_0 (r) \ \varphi_0 (r) \ , \\ \\ r \in [0, \infty )\quad, \quad \varphi_0(0) = 0\ , \ \varphi ' _0(0) = 1 \ . \end{array} \right .\eqno({\rm A})$$

\noi It is assumed that either $V_0$ is positive (repulsive), or else, if it is negative (attractive), it is weak and sustains no bound states. Moreover, we assume
$$\int_0^1 r |V_0(r)|dr < \infty \quad, \quad \int_1^{\infty} r^2 |V_0 (r) |dr < \infty \ . \eqno({\rm B})$$

\noi Under the above conditions on $V_0 (r)$, i.e. absence of bound states and (B), it can be shown that
$$\left \{ \begin{array}{l} \varphi_0 (r) = r + o(r) \ {\rm as} \ r \to 0\ , \  \varphi_0 (r) > 0 \ {\rm for \ all}\ r > 0\ , \\ \\ \varphi_0(r) = Ar + B + o(1)\ {\rm as} \ r\to \infty  \ , \ 0 < A < \infty\ , \ |B| < \infty   \ . \end{array} \right .\eqno({\rm C})$$

A second independent solution of the Schr\"odinger equation is given by
$$\left \{ \begin{array}{l} \chi_0 (r) = \varphi_0 (r) \displaystyle{\int_r^{\infty} {dt \over \varphi_0^2(t)}} \ , \ r > 0 \ ,\\ \\ \chi_0(0) = 1\ , \ W[\varphi_0, \chi_0] = \varphi '_0 \chi_0 - \chi '_0 \varphi_0 = 1   \ . \end{array} \right .\eqno({\rm D})$$

\noi Indeed, it follows from the definition of $\chi_0(r)$ that $\chi ''_0 = V_0 \chi_0$, and $W[\varphi_0, \chi_0] = 1$. Then $\chi_0 (0) = 1$ follows from the first line of (C). One gets then 
$$\left \{ \begin{array}{l} \chi_0 (r) > 0\ {\rm for}\ r \in [0, \infty )\ , \\ \\  \chi_0(r)= \displaystyle{{1 \over A}} + o(1)\qquad {\rm as } \ r \to \infty \ ,   \end{array} \right .\eqno({\rm E})$$

\noi where $A$, strictly positive and finite, is defined in (C).\par

Now, consider the mapping $r \to x(r)$ defined by
$$x(r) = {\varphi_0 (r) \over \chi_0 (r)}\quad , \quad  {dx \over dr} = {\varphi '_0 \chi_0 - \varphi_0 \chi '_0 \over \chi_0^2(r)} = {1 \over \chi_0^2(r)} > 0\ . \eqno({\rm F})$$

It is obvious that the mapping is one-to-one, and is smooth. It is in fact $C^2$ since
$${d^2x \over dr^2} = {-2 \chi '_0 (r) \over \chi_0^3 (r)}\ . \eqno({\rm G})$$

\noi Therefore, we can use $x(r)$ for making a change variable in the Schr\"odinger equation. Note that one has also, according to (C) and (E),
$$\left \{ \begin{array}{l} x(r) = r + o(r) \quad {\rm as}\ r \to 0\ , \\ \\ x(r) = A^2 r + AB + o(1) \qquad {\rm as}\ r \to \infty\ .  \end{array} \right .\eqno({\rm F}')$$

We consider now the equation 
$$\left \{ \begin{array}{l} \varphi '' (r) = V_0 (r) \varphi (r) + [\chi_0 (r)]^{-4} V[x(r)] \varphi (r)\ , \\ \\ \varphi(0) = 0\ , \ \varphi ' (0) = 1 \ .
 \end{array} \right .\eqno({\rm H})$$

\noi Note here that, as we saw before, $\chi_0 (r)$ is a smooth, bounded, and strictly positive function for all $r \geq 0$.

We assume again that 
$$\left \{ \begin{array}{l} \hbox{i) $V_0(r)$ satisfies (B), and sustains no bound states,} \\ \\ \hbox{ii) $V(x)$ satisfies also (B) in the variable $x$,}\\ \\ \hbox{iii) $V(r)$ can have any (finite) number of bound states.}\end{array} \right .\eqno({\rm I})$$

\par \vskip 5 truemm

\noi {\bf Remark 1.} From (B) for $V(x)$, it follows that
$$\int_0^{\infty} x |V(x)|dx < \infty\ . \eqno({\rm J})$$

\noi It then follows from the Bargmann bound for the number $n$ of bound states of $V(x)$ that
$$n(V) \leq \int_0^{\infty} x |V(x)|dx < \infty\ . \ \sq \eqno({\rm K})$$
\par \vskip 5 truemm

If we make now in (H) the change of variable $r \to x(r)$ defined by (F), and the change of function
$$\psi (x) = \left [{\varphi (r) \over \chi_0 (r)}\right ]_{r = r(x)} \ , \eqno({\rm L})$$

\noi $r(x)$ being the inverse mapping, well-defined and also $C^2$, mapping $x \in [0, \infty )$ into $r \in [0, \infty )$, we find
$$\left \{ \begin{array}{l} \ddot{\psi}(x) = V(x) \psi (x) \\ \\ \psi (0) = 0\ , \ \dot{\psi} (0) = 1 \ .\end{array} \right .\eqno({\rm M})$$

\noi Therefore, if (A) and (M) can be solved explicitely for $V_0(r)$ and $V(x)$, then (H) also can be solved explicitely, and we have, according to (L),
$$\varphi (r) = \chi_0 (r)\  \psi (x) = \chi_0 (r)\  \psi \left ( {\varphi_0 (r) \over \chi_0 (r)}\right ) \  . \eqno({\rm N})$$

\noi We have therefore, the following : \par \vskip 5 truemm

\noi {\bf Theorem 1.} Suppose the Schr\"odinger equation $\varphi '' (r) = v(r) \varphi (r)$, $r \in [0, \infty )$, together with $\varphi (0) = 0$, $\varphi '(0) = 1$, can be solved explicitely for two potentials $V_0(r)$ and $V(r)$, both satisfying the integrability conditions shown in (B). We assume i) $V_0$ sustains no bound states~; ii) $V$ can have bound states, their number $n$ being finite according to (K). Then (H) also can be solved explicitely, and its solution is given by (N). This, of course, can be checked directly by differentiating (N). \\

\noi {\bf Remark 2 - Iteration.} Once we have (N), we can start now with $V_0$ and $V_0 + \chi^{-4} (r) V[x(r)]$, and repeat the operation to get an infinite number of potentials.\\

\noi {\bf Generalizations to singular potentials.} Once the theorem is established, it was then shown in ref. $^1$ that one can generalize it to the case where $V_0(r)$ in (H) can be singular and repulsive at the origin, violating therefore $rV_0(r) \in L^1(0, 1)$, or be long range and repulsive at infinity, violating therefore $r^2V_0(r) \in L^1(1, \infty )$, provided always that it sustains no bound states. Also $V(r)$ can be more general than it was assumed. Many explicit examples, illustrating all these cases, were given. Of course, if $V_0$, and or $V$, violate (B) at $r=0$ or $r= \infty$, one must modify, accordingly, the boundary conditions. Full details are given in ref.~$^1$. For each case, one singular example is shown below. One must secure, of course, each time that the corresponding $\chi_0(r)$ does not vanish for $r \geq 0$, i.e. absence of bound states for $V_0$. This is no problem since in all the explicit examples we give for singular potentials, we are dealing with (modified) Bessel functions, and the locations of the zeros of these functions are known.$^7$ One can also introduce angular momentum by adding $\ell (\ell + 1)/r^2$ to $V_0$.$^{1}$ \\

\noi {\bf Bound States.} We have assumed that $V_0$ has no bound states. If $V(x)$ sustains $n$ bound states, then, according to the nodal theorem,$^{5,6}$ $\psi (x)$, solution of (M), has $n$ zeros for $x > 0$. It follows then from (N) that $\varphi (r)$ has the same number of zeros for $r > 0$, and, therefore, that the potential $V_0(r) + \chi_0^{-4} (r)$ $V(x)$ sustains the same number of bound states as $V(r)$.\\

\noi {\bf Regular Examples.} Potentials for which (A) and (M) can be solved explicitely, are many. Not only the classical examples$^{1, 2}$ 
$$\left \{ \begin{array}{l}V(r) = \displaystyle{{\lambda a^2 \over (1 + ar)^4}}  \quad , \quad a >  0 \ , \\ \\ \varphi (r) =\left ( \displaystyle{ {1 + ar \over a \sqrt{\lambda}}}\right ) \sinh \left ( \displaystyle{{\sqrt{\lambda} ar \over 1 + ar }} \right ) \ ; \end{array} \right .
\eqno({\rm O})$$

$$
\left \{ \begin{array}{l}V(r) = \displaystyle{{\lambda b^2 \over (b^2 + r^2)^2}}  \quad , \quad (b >  0) \ , \\ \\ \varphi (r) = \displaystyle{ {(b^2 + r^2)^{1/2} \over  \sqrt{\lambda -1}}} \sinh \left ( \sqrt{\lambda -1} \ Arctg \displaystyle{{r \over b}} \right ) \ ; \end{array} \right .
\eqno({\rm P})$$

\noi and
$$
\left \{ \begin{array}{l}V(r) = \lambda e^{-\mu r}  \quad , \quad \mu  >  0 \ , \\ \\ \varphi (r) = \alpha I_0 \left ( \displaystyle{ {2\sqrt{\lambda} \over  \mu}} \ e^{- \mu r/2} \right ) + \beta K_0  \left ( \displaystyle{{2\sqrt{\lambda} \over \mu}} \ e^{-\mu r/2}   \right ) \ , \end{array} \right .
\eqno({\rm Q})$$

\noi where $I_0$ and $K_0$ are modified Bessel and Hankel functions of order zero,$^7$ and $\alpha$ and $\beta$ are determined as to have $\varphi (0) = 0$, $\varphi ' (0) = 1$, but also the full class of infinite other potentials we found in reference 1, as well as all the Bargmann potentials, etc$^{2,4}$. Therefore, one can use them as $V_0$ and $V$ in (H), and obtain many explicitely soluble examples. The above formulas are written for $\lambda > 1$, or for $\lambda >0$. For $\lambda < 0$, sinh goes to sin, and $I_0$ and $K_0$ to $J_0$ and $N_0$. It is known that, for $\lambda > 0$, $I_0$ and $K_0$ have no zeros for $r \geq 0$,$^7$ and for $\lambda < 0$, Bessel functions have, usually, oscillations. Remember that whatever the potential we choose for $V_0$, it should not sustain bound states, i.e. be either repulsive, or, if attractive, be weak.\\

\noi {\bf Singular Potentials.} One can also include a singular repulsive potential as well (singular at the origin), like 
$$
V_0(r) = {g \over r^n}\quad , \quad g > 0 \quad , \quad n > 2\ .
\eqno({\rm R})$$

\noi And it can be checked easily that everything works as before. We leave the details for the reader. They can be found in $^1$. We just note that here $\chi_0(0) = \infty$, and $\chi_0(\infty ) = 1$. Since $V_0$ is singular now, we must assume $V_0 > 0$, for,  we know that, in such a case, i.e. with singular and attractive potentials at the origin, violating (B), we don't have a unique self-adjoint extension of the Hamiltonian in $L^2 ({I \hskip - 1 truemm R}^3)$.$^{2,3,6,8}$ The simplest case here is to take $n = 4$. Then, the two independent solutions of (A) are given by, up
$$n = 4
\left \{ \begin{array}{l} \chi_0 = \displaystyle{{r \over \sqrt{g}}} \sinh (\sqrt{g}/r)\ , \\ \\ \varphi_0 (r) = r \exp (- \sqrt{g}/r) \ . \end{array} \right .
\eqno({\rm S})$$

\noi Note that, in accordance with $^1$, because the potential violates (B) at the origin, we have $\chi_0 (0) = \infty$, and $\chi_0(\infty ) = 1$. 
As an example of a long range potential, we consider (M) with 
$$
V(x) = {\alpha \over x}\ ,
\eqno({\rm T})$$
\noi and we find, according to $^1$, that the solution is given by 
$$
\psi (x) = \sqrt{{x \over \alpha}} \ I_1\left ( 2 \sqrt{\alpha x}\right ) \ ,
\eqno({\rm U})$$

\noi where $I_1$ is the modified Bessel function.$^7$ If $\alpha > 0$, we would have no bound states. If $\alpha < 0$, $I_1$ goes to $J_1$, the ordinary Bessel function, which has infinitely many oscillations as $x \to \infty$, and, therefore according to the nodal theorem,$^{5,6}$ we would have infinitely many bound states accumulating at energy $E = 0$, as is well-known. It is also known that, for $\alpha > 0$, $I_1$ does not vanish for $x > 0$, and increases exponentially as $x \to \infty$.$^{7}$

Much work has been done, of course, for finding potentials for which the Schr\"o\-dinger equation can be solved. We refer the reader to the paper of G. L\'evai,$^9$ which contains full references to earlier works.\\

\noi {\bf II - A Restricted Class of Potentials}\\

For this purpose, we use a transformation of the Schr\"odinger equation devised some years ago by one of the authors and Harald Grosse, which has very smoothing effects on the potential.$^4$  It is as follows. Consider the radial Schr\"odinger equation at zero energy 
\beq
\label{1e}
\left \{ \begin{array}{l} \varphi '' (r) = V_0 (r) \varphi (r)\ , \quad V_0 \ \hbox{of any sign} \ , \\ \\r \in [0, \infty )\quad , \quad \varphi (0) = 0 \ . \end{array} \right .
\eeq

\noi $V_0(r)$ is assumed to be a regular potential, i.e. to satisfy the Bargmann-Jost-Kohn condition$^{2,3}$
\beq
\label{2e}
rV_0 (r) \in L^1(0, \infty ) \ .
\eeq

\noi We introduce now
\beq
\label{3e}
W_0(r) = - \int_r^{\infty} V_0(t) dt \quad ; \quad U_0 (r) = \int_r^{\infty} W_0 (t) dt \ .
\eeq

\noi Our transformation is now defined by$^4$ 
\beq
\label{4e}
x = \int_0^r e^{2 U_0(t)} dt \quad , \quad \psi (x) = e^{U_0 (r)} \varphi (r) \ .
\eeq

\par \vskip 5 truemm
\noi {\bf Remark 3.} Since, for large values of $r$, we have $|V_0| < r |V_0| \in L^1(\infty )$, $W_0$ is well-defined and is an absolutely continuous function for all $r > 0$. Then it is easily seen that $W_0(r) \in L^1(0, \infty )$. Indeed,
\bea
\label{5e}
|U_0(r)| &=&\left | \int_r^{\infty} W_0(r) dr \right | \leq \int_0^{\infty} \left | W_0 (r)\right | dr \leq \int_0^{\infty} dr \int_r^{\infty} \left | V_0 (t)\right | dt\nn \\ 
&=&\int_0^{\infty} \left | V_0(t) \right | dt \int_0^t dr = \int_0^{\infty} t\left | V_0(t)\right | dt < \infty \ .
\eea

\noi It follows that $U_0(r)$ is a well-defined, bounded, and absolutely continuous function for all $r \geq 0$, and is also continuously differentiable for $r > 0$. $\sq$\\

From the above results, it is obvious that the transformation (\ref{4e}) is a nice smooth transformation, and we have a smooth one to one mapping
\beq
\label{6e}
r \in [0, \infty ) \Leftrightarrow x\in [0, \infty )\ ,
\eeq

\noi and $(dx/dr ) = \exp (2U_0(r)) > 0$. Obviously, we have also
\beq
\label{7e}
\varphi (0) = 0 \Leftrightarrow \psi (0) = 0 \ .
\eeq

\noi Making the transformation (\ref{4e}) in (\ref{1e}), we find, with $\dot{} = {d \over dx}$, 
\beq
\label{8e}
\left \{ \begin{array}{l} \ddot{\psi} (x) = \left [ - W_0^2(r) \ e^{-4U_0(r)}\right ] \psi (x) = \widetilde{V}(x) \psi (x)\\ \\ x \in [0, \infty ) \quad , \quad \psi (0) = 0 \ . \end{array} \right . 
\eeq

\noi Now, since $U_0 (r)$ is a very smooth and bounded function, for all $r \geq 0$, and $U( \infty ) = 0$, it is obvious from the definition (\ref{4e}) that $x$ and $r$ are very close to each other, and we have 
\beq
\label{9e}
\left \{ \begin{array}{ll}  x \simeq e^{2U_0(0)} r + o(r)\quad , \qquad &\hbox{as } r \to 0 \ , \\ \\ x \simeq r + 0(1)\quad , \qquad &\hbox{as } r\to \infty\ . \end{array} \right .
\eeq

\noi Therefore, as far as integrability at $x = 0$ and $x = \infty$ are concerned, we have
\beq
\label{10e}
\int_0^{\infty} x \left [ e^{-4U_0(r)} \ W_0^2(r) \right ] dx \sim \int_0^{\infty} r \ W_0^2 (r) dr \ .
\eeq

\noi To show that the last integral is absolutely convergent is now very easy. Indeed, from the definition of $W_0(r)$, (\ref{3e}), 
\beq
\label{11e}
\left | r\ W_0 (r) \right | \leq \int_r^{\infty} t \left | V_0 (t) \right | dt < \int_0^{\infty} t \left | V_0(t) \right | dt = C < \infty\ .
\eeq 

\noi Therefore, 
\beq
\label{12e}
\left | r\ W_0^2(r)\right | \leq C \left | W_0 (r)\right |\ ,
\eeq

\noi and since $W_0 (r)$ was shown to be $L^1(0, \infty )$, the same is true for $r W_0^2(r)$. Therefore, in (\ref{8e}), $x \widetilde{V}(x) \in L^1(0, \infty )$, and $\widetilde{V}(x)$ is a regular potential.

Consider now the equation
\bea
\label{13e}
\varphi '' (r) &=& \left [ V_0(r) + W_0^2(r) + e^{4U_0(r)}\ V_1(x) \right ] \varphi (r)\nn \\
&=& V(r)\ \varphi (r) \ .
\eea

\noi Note here the variable $x = x(r)$ in $V_1(x)$~! Both $rV_0(r)$ and $xV_1(x)$ are assumed to be $L^1(0, \infty )$, i.e. satisfy (\ref{2e}). Since we showed that $r W_0^2(r)$ was also $L^1(0, \infty )$, and $x$ and $r$ are always of the same order by virtue of (\ref{9e}), the full potential in (\ref{13e}) satisfies the same integrability condition. 

After making the transformation (\ref{4e}), one finds then, with $\dot{} = (d/dx)$,
\beq
\label{14e}
\left \{ \begin{array}{l} \ddot{\psi} (x) = V_1(x)\  \psi (x) \ , \\ \\ x \in [0, \infty ) \quad , \quad \psi (0) = 0 \quad , \quad xV_1(x) \in L^1(0, \infty ) \ .\end{array} \right .
\eeq

\noi Therefore, if the equation (\ref{14e}) with $V_1(x)$ can be solved explicitely, we can solve also explicitely (\ref{13e}), and its solution is given by 
\beq
\label{15e}
\varphi (r) = e^{-U_0 (r)} \ \psi (x) 
\eeq

\noi where $x$ is defined in (\ref{4e}). Therefore, we have completed our programme, i.e. starting from two potentials $V_0$ and $V_1$,  and knowing that the Schr\"odinger equation can be solved explicitely for $V_1$, to find a new potential for which the same holds. One can check again, directly, that (\ref{15e}) is, indeed, a solution of (\ref{13e}).

The connection with ref. $^1$ is as follows. Consider
\beq
\label{16e}
\psi '' _0= \left ( V_0 + W_0^2\right ) \psi_0 \ .
\eeq

\noi A solution of this equation, called $\chi_0 (r)$,  with $\chi_0(0) = 1$, is
\beq
\label{17e}
\chi_0 (r) = {e^{-U_0 (r)} \over e^{-U_0(0)}}\ .
\eeq

\noi This has no zeros for $r \geq 0$. Therefore, the physical solution of (\ref{16e}), called $\varphi_0(r)$, with $\varphi_0(0) = 0$, and given by$^1$
\beq
\label{18e}
\varphi_0 (r) = \chi_0 (r) \int_0^r {dt \over \chi_à^2(t)}\ ,
\eeq

\noi as can be checked easily, has also no zeros for $r > 0$. Therefore, according to the nodal theorem$^{5, 6}$, $V_0 + W_0^2$ cannot have bound states, whatever the sign of $V_0$ is. According to (\ref{17e}), the full potential in (\ref{13e}) can be written
\beq
\label{19e}
V(r) = \left ( V_0(r) + W_0^2 (r) \right ) + \lambda \chi_0^{-4}(r) \ V_1(x)\ .
\eeq

\noi This is in complete analogy with reference $^1$, where we had
\beq
\label{20e}
V(r) = V_0(r) + \lambda \chi_0^{-4}(r)\ V_1(x)\ ,
\eeq

\noi assuming there $V_0(r)$ to be either positive, or negative but weak enough in order not to have bound states in (\ref{1e}). In the present paper, $V_0$ is replaced by $V_0 + W_0^2$, with the same properties, but now with simple explicit solutions given by (\ref{17e}) and (\ref{18e}), and no restrictions on the sign of $V_0$.\\

\noi {\bf Bound States.} So far, we did not assume anything on the signs of $V_0$ and $V_1$ in (\ref{13e}). They can have any sign. As we showed, $V_0 + W_0^2$ alone cannot have bound states. So in order to have bound states, $V_1(x)$ must be negative, and strong enough. Now, again, the nodal theorem$^{5, 6}$, applied to (\ref{14e}), shows that, if $V_1(x)$ sustains $n$ bound states, then $\psi (x)$ has $n$ zeros (nodes) for $x > 0$. Therefore, according to (\ref{15e}), $\varphi (r)$ also has $n$ nodes for $r > 0$. So, there are $n$ bound states also for the full potential (\ref{18e}). In conclusion, whatever the sign of $V_0$ is, $(V_0 + W_0^2)$ sustains no bound states, and $V_0 + W_0^2 + \chi_0^{-4} V(x)$ and $V(r)$ have the same number of bound states, where $x$ is given in (\ref{4e}).\\

\noi {\bf Higher $\ell$.} Generalization to include the centrifugal potential $\ell (\ell + 1)/r^2$ in $V_0 + W_0^2$ is straightforward.$^4$. We define now
\beq
\label{24e}
\left \{ \begin{array}{l}W_{\ell} (r) = - \displaystyle{\int_r^{\infty}} V_0(t) \ t^{-2\ell}\ dt \ , \\ \\ U_{\ell} (r) = \displaystyle{\int_r^{\infty}} W_{\ell} (t) \ r^{2\ell} \ dt  \ . \end{array} \right .
\eeq

\noi Since
\beq
\label{25e}
\left | r^{2 \ell} W_{\ell} (r) \right | \leq \int_r^{\infty} V_0 (t) \ dt < \infty\quad, \quad \forall\ r > 0\ ,
\eeq

\noi it is obvious that $r^{2\ell} W_{\ell} (r)$, like $W_0(r)$, is bounded for $r > 0$, and $\in L^1 (0 , \infty )$. And $U_{\ell}(r)$ is a nice bounded and smooth function for all $r \geq 0$, as was the case for $U_0(r)$. Now, we define,
\beq
\label{26e}
\chi_{\ell} (r) = r^{-\ell} \ e^{-U_{\ell} (r)}\ .
\eeq

\noi It is then easily seen that $\chi_{\ell} (r)$ satisfies
\beq
\label{27e}
\chi ''_{\ell}(r) = {\ell (\ell + 1) \over r^2} \ \chi_{\ell}(r)+ V_0\ \chi_{\ell} + r^{4\ell} \ W_{\ell}^2 (r) \chi_{\ell} (r)\ .
\eeq

\noi The generalization of (\ref{13e}) is now 
\beq
\label{28e}
\varphi ''_{\ell} (r) = \left [ V_0(r) + r^{4\ell}\ W_{\ell}^2(r) + {\ell (\ell + 1) \over r^2} + r^{4\ell} \ e^{4U_{\ell}}\  V_1(x) \right ] \varphi_{\ell} (r) \ , 
\eeq

\noi and one finds, after the transformations
\beq
\label{29e}
x = x(r) = \int_0^r t^{2\ell}\ e^{2U_{\ell}(t)}\ dt \quad , \quad \psi_{\ell} (x) = r^{\ell}\ e^{U_{\ell} (r)}\ \varphi_{\ell} (r) \ ,
\eeq

\noi the differential equation
\beq
\label{30e}
\ddot{\psi}_{\ell} (x) = V_1(x) \ \psi_{\ell} (x) \ .
\eeq

\noi Therefore, the conclusion is exactly as for the case $\ell = 0$. Remember that we have assumed $xV_1(x) \in L^1(0, \infty )$. The solution of (\ref{28e}) is, therefore, given by 
$$\varphi_{\ell}(r) = r^{-\ell} \ e^{-U_{\ell}(r)} \ \psi_{\ell} (x) \ . \eqno(25')$$

\noi Making $\ell = 0$, we find, of course, (\ref{13e})-(\ref{15e}).\\

\noi {\bf Examples.} With the potentials (O)-(Q) for $V_0$ and $V_1$, and (\ref{3e}), (\ref{13e}), (\ref{14e}) and (\ref{15e}), one can construct easily many explicit examples. We leave the details to the reader. With (R) for $V_0$, $n=4$, and (\ref{17e}) and (\ref{18e}), it is easily found that
$$\chi_0 (r) = \exp (g/6r^2)\ , \eqno(17')$$

\noi and
$$\varphi_0 (r) = \left \{ \begin{array}{l} r^3 \exp (-g/6r^2) + \cdots\ , \quad r\to 0\\ \\ r - \sqrt{g} + \cdots\ , \quad r \to \infty  \ .\end{array} \right .  \eqno(18')$$

\noi One can then take for $V_1$ one of the potentials (O)-(Q).\\

\noi {\bf III -  The General Case} \\

In this section, we are going to combine both transformations, namely, the transformation of section I, and the tranformation of section II. Consider the three potentials
\beq
\label{36e}
\left \{ \begin{array}{l} \hbox{$V_0(r)$ satisfying (B) : $rV_0(r) \in L^1(0,1)$, $r^2V_0(r)\in L^1(1, \infty)$,}\cr \hbox{and no bound states~;}\cr\cr \hbox{$V_1(r)$ satisfying $rV_1(r) \in L^1(0, \infty )$~;}\cr\cr \hbox{$V(r)$ satisfying $rV(r) \in L^1(0, \infty )$.}\cr\cr\hbox{$V_1(r)$ and $V(r)$ can have bound states.}\end{array} \right .
\eeq

We assume that one can solve explicitely
\beq
\label{37e}
\left \{ \begin{array}{l} \varphi '' (r) = v(r) \ \varphi (r)\quad, \quad v = V_0 (r)\ , \ V(r)\ ,\cr\cr \varphi (0) = 0 \quad , \quad  \varphi ' (0) = 1 \ . \end{array} \right .
\eeq

\noi Nothing is assumed for the explicit solution for $v = V_1(r)$.

As we saw in section I, the assumptions on $V_0(r)$, entail the existence of the solution $\chi_0(r)$ of (\ref{37e}), given by (D), such that
\beq
\label{38e}
\left \{ \begin{array}{l} \chi_0(r) > 0\ {\rm for \ all}\ r \geq 0 \ , \cr \cr \chi_0 (0) = 1\ , \ \chi_0(\infty) = \displaystyle{{1 \over A}}\ , \ 0 < A < \infty\ , \end{array} \right .
\eeq

\noi that is, a smooth and strictly positive bounded function. We introduce now
\beq
\label{39e}
\left \{ \begin{array}{l}W_1(r) = - \displaystyle{\int_r^{\infty}} V_1(t) \ \chi_0^2(t)\ dt \ , \cr \cr U_1(r) = \displaystyle{\int_r^{\infty}} W_1(t)  \displaystyle{{dt \over \chi_0^2(t)}}\ .  \end{array} \right .
\eeq

\noi It follows now that $W_1$ and $U_1$ have the same properties as $W$ and $U$ introduced earlier in (\ref{3e}), and shown in {\bf Remark 1}, (\ref{5e}), and (\ref{12e})~:
\beq
\label{40e}
\left \{ \begin{array}{l} W_1(r)\in L^1(0, \infty),\cr \cr \hbox{$U_1(r)$ smooth and bounded for all $r\geq 0$,}\cr\cr rW_1^2(r) \in L^1(0, \infty ).\end{array} \right .
\eeq
\noi Let us now introduce, as in section II, the mapping
\beq
\label{41e}
r \to x = x(r) = \int_0^r e^{2U_1(t)} {dt \over \chi_0^2(t)}\ .
\eeq

\noi It is a smooth, and twice differentiable one-to-one mapping for $r > 0$~:
\beq
\label{42e}
\left \{ \begin{array}{l} r\in [0, \infty ) \Leftrightarrow x \in [0, \infty ) \ ,\cr \cr r = 0 \Leftrightarrow x= 0\quad , \quad r = \infty \Leftrightarrow x = \infty \ .
\end{array} \right .
\eeq

\noi Indeed, $U_1(\infty ) = 0$, $\chi_1(\infty ) = {1 \over A} \not= \infty$, as shown in (E), and
\beq
\label{43e}
\left \{ \begin{array}{l}\displaystyle{{dx \over dr}} =  \displaystyle{{e^{2U_1(r)} \over \chi_0^2 (r)}} > 0\quad , \quad \to A^2\ {\rm for}\ r \to \infty \ ,\cr \cr \displaystyle{{d^2x \over dr^2}} = - \left [ \displaystyle{{2W_1(r) \over \chi_0^4(r)}} + \displaystyle{{2\chi '_0(r) \over \chi_0^3(r)}}\right ] e^{2U_1(r)}\  \hbox{bounded for all $r > 0$}\ .
\end{array} \right .
\eeq

\noi Remember that $\chi ''_0 = V_0 \chi_0$. $\chi_0$ being continuous and bounded for all $r\geq 0$, and $V_0 \in L^1$ by assymption for all $r > 0$, the same is true for $\chi ''_0$~: $\chi ''_0(r) \in L^1$ for $r > 0$. Therefore, $\chi '_0(r)$ is bounded and continuous for $r > 0$. The inverse mapping $r(x)$ is also, of course, a smooth and twice differentiable mapping for $x > 0$. Also, it is obvious from the first part of (\ref{43e}), that
\beq
\label{44e}
\left \{ \begin{array}{l}x(r) = e^{2U_1(0)} r + o(1)\quad , \quad r \to 0\ ,\cr \cr x(r) = A^2r + o(r)\quad , \quad r \to \infty\ .
\end{array} \right .
\eeq

\noi $x$ and $r$ are, therefore, of the same order as $r \to 0$ or $r\to \infty$.

Consider now the Schr\"odinger equation at zero energy
\beq
\label{45e}
\left \{ \begin{array}{l}\varphi '' (r) = \left [ V_0(r) + V_1(r) + \chi_0^{-4}(r) W_1^2(r) + \chi_0^{-4} (r) e^{4U_1(r)} V(x) \right ] \varphi (r)\ , \cr \cr \varphi (0) = 0\ .
\end{array} \right .
\eeq

 Note here, again, the appearance of $x$, given by (\ref{41e}), in $V(x)$. Remember also that $rW_1^2(r) \in L^1(0, \infty )$, as shown in (\ref{40e}). It follows that all the potentials in (\ref{45e}) are regular, i.e. $rv(r) \in L^1(0, \infty )$ by assumption. Making the change of function, where $U_1$ is defined by (\ref{39e}),
\beq
\label{46e}
\left \{ \begin{array}{l}\varphi (r) \to \psi (x) = \left [ \displaystyle{{\varphi (r) e^{U_1(r)} \over \chi_0 (r)}}\right ]_{r=r(x)}\ , \cr \cr \varphi (0) = 0\Leftrightarrow \psi (0) = 0\ , \hbox{which is obvious,}
\end{array} \right .
\eeq

\noi differentiating twice $\psi (x)$ with respect to $x$ defined by (\ref{41e}), and using the first part of (\ref{43e}) and (\ref{45e}), we find
\beq
\label{47e}
\left \{ \begin{array}{l}\ddot{\psi} (x) = V(x) \ \psi(x)\ .\cr \cr \psi (0) = 0\ .
\end{array} \right .
\eeq
 
 \noi Since it was assumed that this equation can be solved explicitely, we have achieved our goal, and the solution of (\ref{45e}), according to (\ref{46e}), is given by
 \beq
 \label{48e}
 \varphi (r) = \chi_0 (r) \ e^{-U_1(r)}\ \psi (x) \ ,
 \eeq
 
 \noi where $x$ is explicitely defined by (\ref{41e}) in terms of $\chi_0 (r)$ and $V_1(r)$ through (\ref{39e}). We can summarize our results in the following theorem~:\\
 
 \noi {\bf Theorem 2.} Given three potentials $V_0$, $V_1$, and $V$ satisfying the assumptions (\ref{36e}), and assuming that (\ref{37e}) can be solved explicitely for $V_0$ and $V$, the solution of (\ref{45e}) is given explicitely by (\ref{48e}). Obviously, we have $\varphi  (0) = 0$.\\

\noi {\bf Final Remarks.} Making $V_0(r) = 0$, i.e. $\chi_0 (r) = 1$, we find the results of section II, and making $V_1(r) = 0$, the results of ref. $^1$, as given in the introduction. Also, having now a new potential with the explicit solution $\varphi$, we can repeat the operation with $V_0$ and the new potential, and continue indefinitely the process. We should note here that all the explicit examples of potentials we have given in ref. $^1$, some of which are reproduced in the present paper, lead to soluble Schr\"odinger equation for any coupling constant in front of the potential. We have, therefore, a great (infinite) variety of soluble potentials. Also, as we said in the Abstract, once the solutions are exhibited, one can check directly, by differentiation, that they satisfy indeed the appropriate equations. One can, of course, include also here the angular momentum, either by proceeding as in section II, or else by making
\beq
\label{41NEW}
\left \{ \begin{array}{l} V_0 (r) \to V_0 (r) + \displaystyle{{\ell ( \ell + 1) \over r^2}} \ ,  \\ \\ r^{2\ell+2} \ V_0 (r) \in L^1(1 , \infty ) \end{array}\right .
\eeq\

\noi in (\ref{37e}), and replacing $\varphi_0 (r)$ and $\chi_0 (r)$ by the appropriate solutions of $\varphi_{\ell} (r)$ and $\chi_{\ell} (r)$. Details are given in .$^{1}$ Since $V_0$ is assumed to sustain no bound states, the same is true when one makes (\ref{41NEW}). And one shows again that the mapping $r \to x(r)$ is one to one, and twice differentiable. One has $r \in [0, \infty ) \Leftrightarrow x \in [0, \infty )$, and one can proceed as shown before in this section by replacing $\chi_0 (r)$ by $\chi_{\ell} (r)$ in (\ref{39e}), (\ref{41e}), (\ref{45e}), and (\ref{46e}). We obtain now another kind of potentials with explicit solutions.\\

\noi {\bf Acknowledgments.} One of the authors (KC) would like to thank Professors Kenro Furutani and Takao Kobayashi, and the Department of Mathematics of the Science University of Tokyo, for warm hospitality and financial support.\\

\noi {\bf References}

\begin{enumerate}
\item K. Chadan and R. Kobayashi (2005), ArXives math-ph/0510047, 12 Oct. 2005. Submitted to J. Phys. A. In the figure 2, the curves I and II should be interchanged.
\item R. Newton, Scattering Theory of Waves and Particles (Springer, New York, 2nd edition, 1982). See specially chapter 14 for many soluble examples.
\item A. Galindo and P. Pascual, Quantum Mechanics, two volumes (Springer, Berlin, 1990), vol. I.
\item K. Chadan and H. Grosse, J. Phys. A, {\bf 16}, 955 (1983).
\item R. Courant and D. Hilbert, Methods of Mathematical Physics (Intescience, New York, 1952), volume I.
\item E.-A. Coddington and N. Levinson, Theory of Ordinarly Differential Equations (Mc-Graw-Hill, New York, 1955).
\item A. Erd\'elyi, editor, Higher Transcendal Functions, vol. II (McGraw-Hill, New York, 1953).
\item E. Hille, Lectures on Ordinary Differential Equations (Addison-Wesley, Reading, 1969).
\item G. L\'evai, J. Phys. A, {\bf 22}, 689 (1984).
\end{enumerate}

\end{document}